\documentclass[a4paper]{jpconf}
\usepackage{graphicx}
\begin{document}
\title{Physics performance studies for the ALICE inner tracker upgrade}
\author{Johannes Stiller for the ALICE Collaboration}
\address{Physikalisches Institut, University of Heidelberg, Im Neuenheimer Feld 226, 69120 Heidelberg}
\ead{stiller@physi.uni-heidelberg.de}
\begin{abstract}
During the second long shutdown of the LHC in 2018, the ALICE Collaboration plans to install an upgrade of the ALICE Inner Tracking System (ITS) in the central barrel with seven layers of silicon detectors starting at 2.2 cm radial distance from the interaction region and a material budget as low as 0.3 \% radiation length per layer. A single-hit resolution of $\mathrm{4~\mu m}$ and a readout rate capability of up to 50 kHz in Pb--Pb collisions will allow new and unique measurements in the heavy-quark sector, i.e. charm and beauty. Using detailed Monte Carlo simulations of pp and Pb--Pb collisions we study the performance for heavy-flavor detection with an upgraded ITS in the following benchmark analyses: Charm meson and baryon production, i.e. $\mathrm{D^{0} \rightarrow K^{-}\pi^{+}}$ and $\mathrm{\Lambda_{c}^{+} \rightarrow pK^{-}\pi^{+}}$, and beauty meson and baryon production, i.e. displaced vertices of $\mathrm{B^{+} \rightarrow \overline{D}^{0}\pi^{+}}$ and $\mathrm{\Lambda_{b} \rightarrow \Lambda_{c}^{+}\pi^{-}}$.
\end{abstract}
\section{Introduction}
A Large Ion Collider Experiment (ALICE)~\cite{ALICE} is designed to study strongly interacting matter at extreme energy densities in high-energy nucleus-nucleus (AA) collisions at the CERN Large Hadron Collider (LHC). The individual subsystems of the detector are arranged in a central barrel and a forward muon spectrometer, which are specifically designed to study Pb--Pb as well as pp and p-Pb collisions as reference. The main tracking devices in the central barrel are, in the order from closest to furthest away from the interaction region, the Inner Tracking System (ITS), the Time Projection Chamber (TPC), the Transition Radiation Detector (TRD) and the Time Of Flight (TOF) system. At central rapidity and down to low transverse momenta ($p_{\rm T}$), ALICE has unique particle identification (PID) capabilities as well as high tracking and vertexing precision. These allow for a detailed characterization of the Quark-Gluon Plasma (QGP). Within this scope the light quark sector is extensively studied by ALICE. For example, the charged particle production in Pb--Pb collisions was determined via the nuclear modification factor ($\mathrm{R_{AA}}$) and it was observed that the suppression of high-$p_{\rm T}$ particles strongly depends on event centrality~\cite{RaaLight}. For the first time, higher harmonic anisotropic flow was measured for charged particles in Pb--Pb collisions~\cite{LightFlow}, giving insight into initial spatial anisotropy and fluctuations. Further, a large enhancement in the strange baryon/meson-ratio was observed~\cite{Enhancement}, which may be explained by coalescence models. In addition, ALICE has performed several measurements in the heavy-flavor sector. A strong suppression of $\mathrm{D}$ meson production in Pb--Pb at LHC energies was measured for the first time down to $p_{\rm T}>\mathrm{2~GeV/}c$~\cite{Raa}. In comparison, models predict a hierarchy in the energy loss, $\mathrm{\Delta E_{g}>\Delta E_{c}>\Delta E_{b}}$, for example due to the dead cone effect~\cite{deadCone}. Furthermore, data from Pb--Pb collisions indicates a non-zero elliptic flow of $\mathrm{D}$ mesons~\cite{flow}, which suggests that charm quarks might participate in the collective expansion of the system. However, to achieve deeper insight, especially within the scope of rare heavy quark production, high-statistics, precision measurements at very low $p_{\rm T}$ are required.
\section{Future LHC and ALICE upgrade strategy}
The so-called \it Phase 0 \rm of the LHC schedule was completed in the beginning of 2013. After the first long shutdown from 2013 to 2014, in which the ALICE detector completion is scheduled, the \it Phase 1 \rm program will then start in 2015 until 2017. In this time period, a total of 1~$\mathrm{nb^{-1}}$ of Pb--Pb at $\mathrm{\sqrt{s_{NN}}}$ = 5.5~TeV and further pp and p--Pb reference data are planned to be collected. In the second long shutdown from 2017 until 2018, significant detector upgrades are planned. Beginning in 2018 with the \it \mbox{Phase 2}, \rm the LHC will increase its luminosity with Pb beams up to an interaction rate of about 50~kHz, which corresponds to an instantaneous luminosity of $\mathrm{6 \times 10^{27}\rm~cm^{-2}s^{-1}}$. With the high-rate upgrade a total amount of data of more than 10~nb$^{-1}$ (Pb--Pb) and more than 6~$\mathrm{pb^{-1}}$ (pp reference) will be available in ALICE. The ALICE upgrade concept~\cite{ALICEloi} aims for new high-precision AA measurements of rare probes at low $p_{\rm T}$. Among these are measurements of $\mathrm{D}$ mesons to zero $p_{\rm T}$, charm and beauty baryons (accessible for the first time) and the measurement of beauty via displaced $\mathrm{D^{0}\rightarrow K\pi}$ (accessible for the first time) to almost zero $p_{\rm T}$. The latter is limited by the statistical uncertainty obtained, which is much larger than the systematic uncertainty from theory-driven methods. Within this concept, the ALICE upgrade entails a new, high-resolution and low-material ITS. This detector yields significant improvements for vertexing and tracking at low $p_{\rm T}$, namely of the pointing resolution (up to a factor of 5 better) as well as the tracking efficiency and the momentum resolution. Further, for low-$p_{\rm T}$ heavy-flavor measurements, a continuous readout is planned at a rate of up to 50~kHz for Pb--Pb and several MHz for pp collisions, with High Level Trigger data compression based on charge clusters stemming from particle tracks. 
This is necessary because of the very high trigger rates for the most interesting signals, as background candidates that pass the selections would also fire the trigger. In case of a conventional trigger and due to the small signal-over-background ratio (S/B) of many signals, a minimum $p_{\rm T}$ threshold would be required to stay at affordable rates, i.e. for $\mathrm{\Lambda_{c}^{+}}$ with $p_{\rm T}~\mathrm{>2~GeV/}c$ gives a rate of 16~kHz. Such a cut at trigger level prevents the measurement at very low $p_{\rm T}$. Besides a new ITS readout, an upgrade of the readout systems of most of the detectors of the central barrel to a pipelined readout is foreseen, including a complete replacement of the TPC readout chambers with gas electron multipliers (GEMs) at the endcaps.
\section{ITS upgrade concept}
\label{ITSupgrade}
The rareness and the low S/B ratio of the desired signals at low $p_{\rm T}$ pose strong constraints on the design of the upgraded ITS~\cite{ITScdr}. 
In order to achieve the required improvement of the impact parameter resolution, the radial position of the first layers of silicon detectors will be moved closer to the interaction point ($\mathrm{39~mm \rightarrow 22~mm}$). Depending on the final design, the total material budget $\mathrm{X/X_{0}}$ per layer will be reduced from presently 1.14~\% to as low as 0.3~\%. 
The overall pixel size improves from currently $\mathrm{50~\mu}$m (r$\mathrm{\phi}$) $\mathrm{\times~425~\mu}$m (z) to $O\mathrm{(20~\mu}$m $\mathrm{\times ~20~\mu}$m) or $O\mathrm{(50~\mu}$m $\mathrm{\times~50~\mu}$m) for monotlithic and state-of-the-art hybrid pixels respectively. An improved tracking efficiency and $p_{\rm T}$ resolution at low $p_{\rm T}$ is achieved by adding an additional detection layer (7 in total) and increasing the pixel granularity. The radial extension increases to \mbox{$22-430$~mm}. In addition it will be possible to quickly insert or remove the detector for yearly maintenance. Two design options have been studied for the upgraded ITS, either consisting of seven layers of pixel detectors or of three inner layers of pixel detectors and four outer layers of strip detectors. The first option yields better standalone tracking efficiency and $p_{\rm T}$ resolution, whereas the second design has better standalone PID capabilities. Here, only the first option will be addressed.
\section{Physics performance of the upgraded ITS}
The reconstruction of $\mathrm{D^{0}\rightarrow K\pi}$ decays, which was performed on Pb--Pb data with good precision, is used as a benchmark to quantify the improved performance of the new ITS. In order to estimate the ITS upgrade performance, a so-called \it Hybrid \rm simulation approach was used. This is a simple scaling of the residuals of impact parameters ($\mathrm{d_{0,r\phi},\ d_{0,z}}$) and transverse momentum based on their true values from Monte Carlo (MC) simulations. The scaling factors are the corresponding ratios of the "upgrade to current" resolution for these parameters. 
\begin{figure}[h]
\begin{minipage}[t]{18pc}
\includegraphics[height=0.75\textwidth]{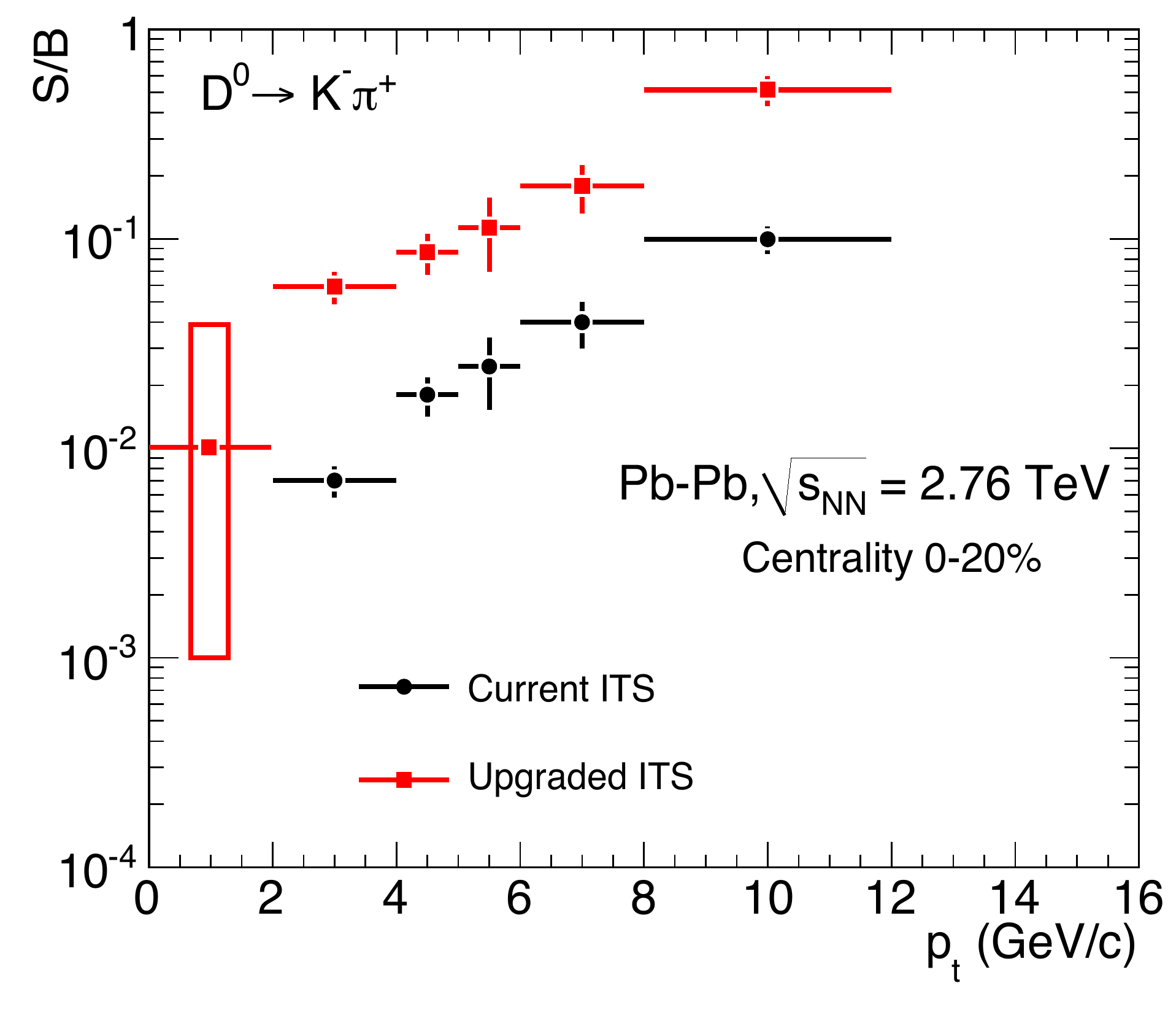}
\caption{\label{DSoverB}Comparison of the $\mathrm{D^{0}}$ signal-over-background ratio for current and upgraded ITS~\cite{ITScdr}.}
\end{minipage}\hspace{1.7pc}%
\vspace{0pc}
\begin{minipage}[t]{18pc}
\includegraphics[height=0.75\textwidth]{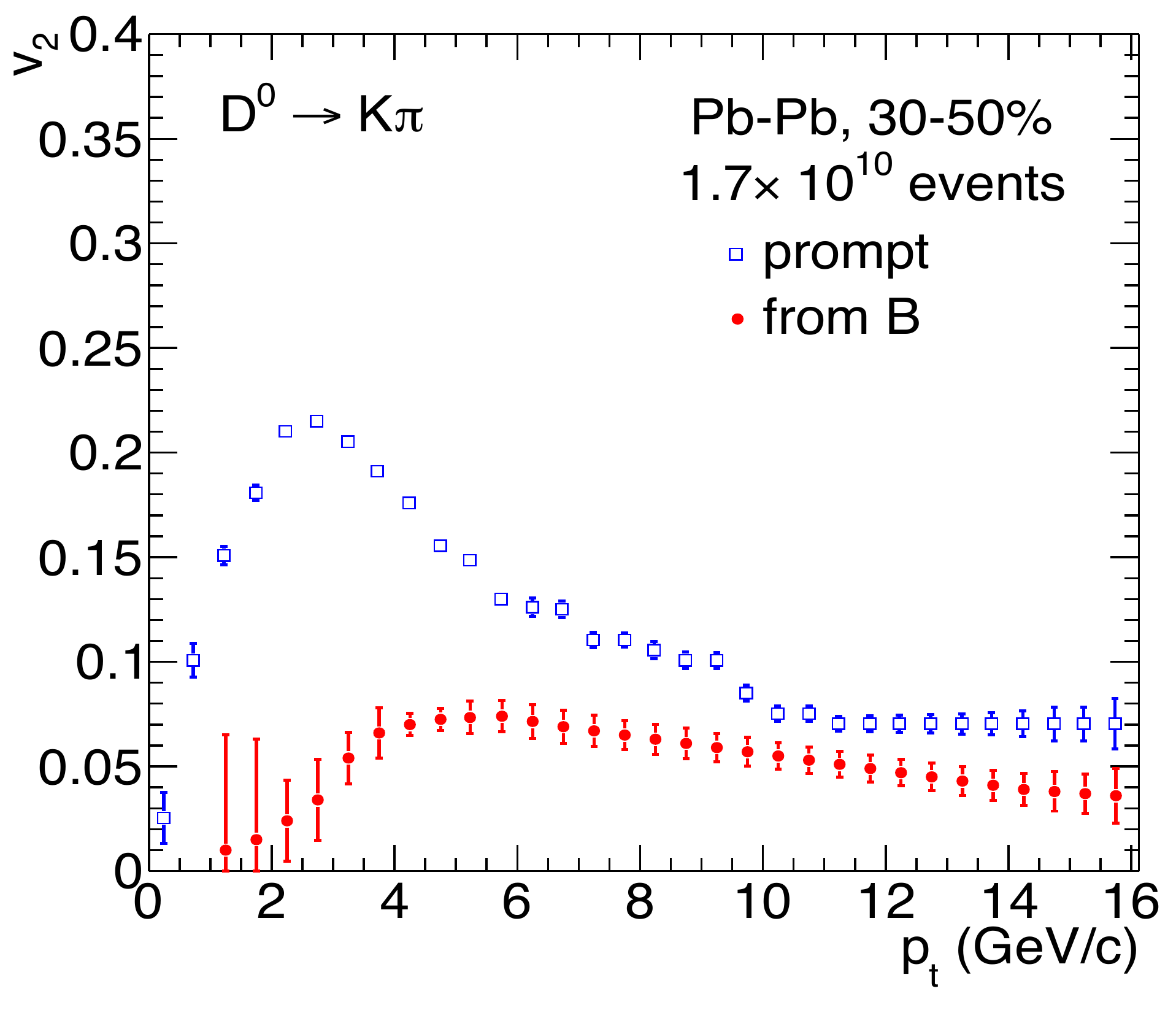}
\caption{\label{Dv2}Estimated statistical uncertainty on $\mathrm{v_{2}}$ of prompt and secondary $\mathrm{D^{0}}$ mesons for $\mathrm{L_{int}=10~nb^{-1}}$~\cite{ITScdr}.}
\end{minipage} 
\end{figure}
\vspace{-2ex}

\begin{figure}[h]
\begin{minipage}{18pc}
\centering
\includegraphics[height=.76\textwidth]{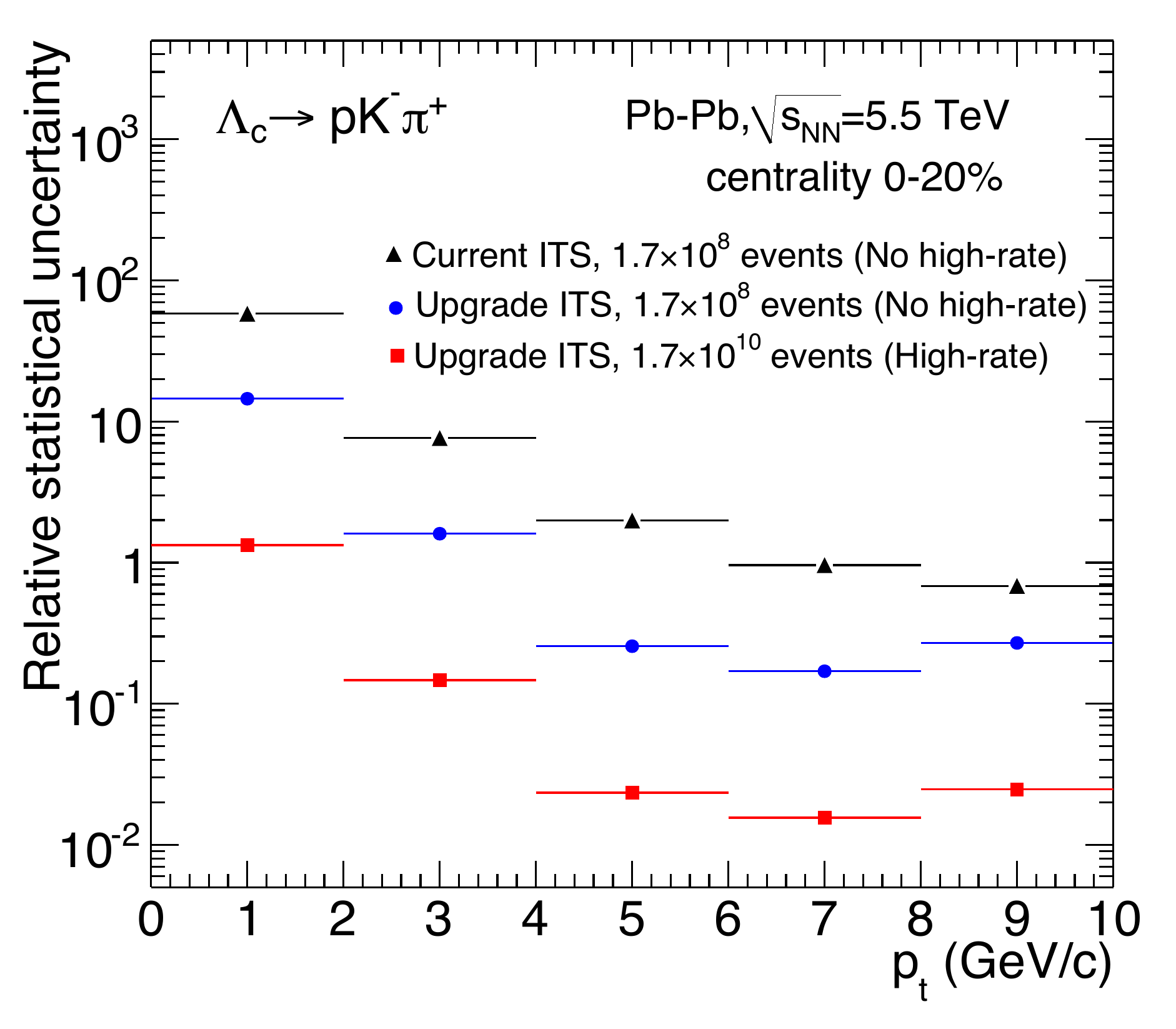}
\caption{\label{LSoverB}Comparison of statistical precision for different cases of ITS performance and integrated luminosity~\cite{ITScdr}.}
\end{minipage}\hspace{1.7pc}%
\vspace{0pc}
\begin{minipage}{18pc}
\centering
\includegraphics[height=.77\textwidth]{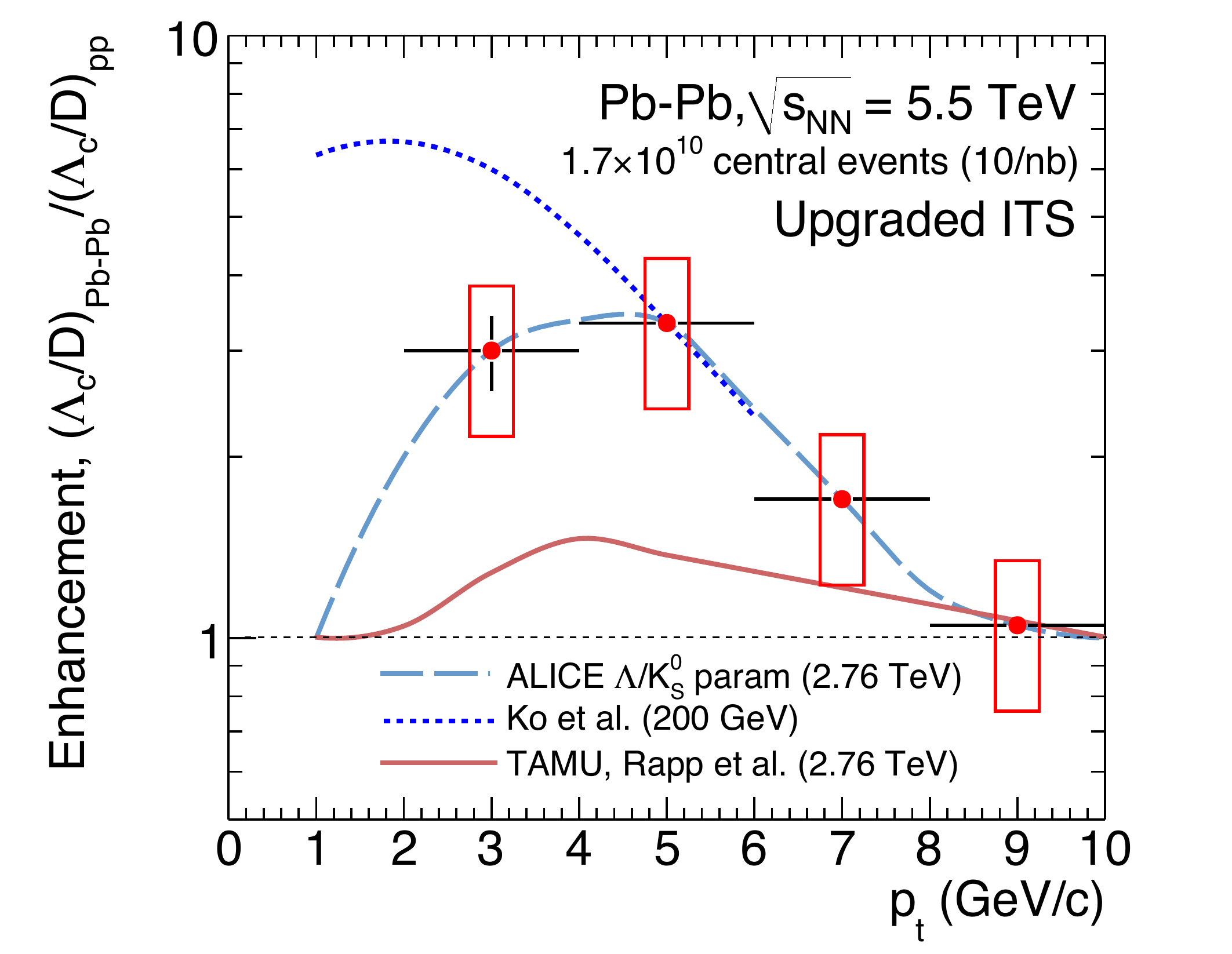}
\caption{\label{LambdaCmodel}Enhancement of the $\Lambda_{c}^{+}/D$ ratio in central Pb--Pb with respect to pp collisions for $\mathrm{L_{int}=10~nb^{-1}}$~\cite{ITScdr}.}
\end{minipage} 
\end{figure}

As visible in Fig.~\ref{DSoverB}, the S/B ratio of the measurement improves substantially due to the improved separation of the signal topology from the background. In both scenarios, the same selection cuts were applied and a similar amount of signal was selected. In the range $\mathrm{0}<p_{\rm T}<\mathrm{2~GeV/}c$ the expected signal and background yields were scaled from pp FONLL predictions at \mbox{2.76~TeV}, as the D meson productions have never been measured in central Pb--Pb collisions in this $p_{\rm T}$ range. The spread of the $\mathrm{R_{AA}}$ (varied between 0.3 and 1) and of the FONLL prediction are the major sources of the rather large uncertainty. The cut efficiencies were determined using MC simulations. Scaling to an integrated luminosity of $\mathrm{10~nb^{-1}}$, a statistical significance above one hundred is measured at any $p_{\rm T}$, which leads to higher precision and extended $p_{\rm T}$ range of the measurements of $\mathrm{v_{2}}$ and $\mathrm{R_{AA}}$. The former is shown in Fig.~\ref{Dv2} for prompt $\mathrm{D}$ and secondary $\mathrm{D}$ from $\mathrm{B}$, whose relative contributions are determined from the analysis of $\mathrm{D^{0}}$ displacement from the primary vertex. Here, $\mathrm{v_{2}}$ was calculated based on the in-plane and out-of-plane yields with respect to the Event Plane (EP) direction, neglecting higher harmonics. Only statistical uncertainties are shown, as most of the systematic uncertainties cancel in the ratio. Further, first performance studies on the $\mathrm{D^{+}_{s}}$ production in Pb--Pb collisions with the upgrade ITS show, that the measurements of $\mathrm{R_{AA}}$ and $\mathrm{v_{2}}$ would largely benefit from the improved tracking precision and high statistics provided by the \it high-rate \rm upgrade of the ALICE central barrel. Another benchmark case is the measurement of charmed baryons ($\mathrm{\Lambda_{c}^{+}\rightarrow p K^{-} \pi^{+}}$), which strongly relies on the excellent ALICE PID capabilities with TPC and TOF. However, a $\Lambda_{c}^{+}$ signal was not observed in Pb--Pb data due to the large combinatorial background. Here, because of the short mean proper decay length of the $\mathrm{\Lambda_{c}^{+}}$, $\mathrm{c\tau \approx 60~\mu m}$, very high spatial resolution is needed in order to cleanly identify the secondary vertex. With the improved detector the most effective cut variables are the $\mathrm{\cos{(\theta_{p})}}$, the decay length and the requirement of a minimum $p_{\rm T}$ of the three decay tracks. The pointing angle $\mathrm{\theta_{p}}$ describes the angle between the straight connection line of primary and secondary vertex and the direction of the reconstructed momentum vector of the decaying particle. As shown in Fig.~\ref{LSoverB}, the relative statistical uncertainty improves significantly for all $p_{\rm T}$. Here, three different scenarios are considered, namely the current and upgraded ITS with \it no high-rate \rm capability ($\mathrm{0.1~nb^{-1}}$) and the upgraded ITS with \it high-rate \rm capability ($\mathrm{10~nb^{-1}}$). In the \it high-rate \rm scenario, $\mathrm{\Lambda_{c}}$ production should be measurable down to $p_{\rm T}=\mathrm{2~GeV/}c$. Based on this new measurement, for example the enhancement $\mathrm{(\Lambda_{c}^{+}/D)_{\rm Pb-Pb}/(\Lambda_{c}^{+}/D)_{\rm pp}}$ can be determined, as shown in Fig.~\ref{LambdaCmodel}. The points are drawn on a line (dashed), which captures the trend and magnitude of the $\mathrm{\Lambda /K^{0}_{S}}$ ratio. Systematic uncertainties, which mainly come from feed-down from  $\mathrm{\Lambda_{b}}$ decays, are displayed as boxes. Two model calculations are presented to indicate the expected sensitivity of the baryon to meson ratio in the measurement. The tagging of $\mathrm{\Lambda_{c}^{+}}$ also allows for the full reconstruction of $\mathrm{\Lambda_{b} \rightarrow \Lambda_{c}^{+}\pi^{-}}$. 
\vspace{-2ex}
\section{Conclusion and outlook}
The ALICE experiment shows outstanding performance in studying strongly interacting matter. For a deeper understanding of, e.g., heavy-flavor energy loss mechanisms, azimuthal anisotropy and in-medium hadronization, the ALICE upgrade concept foresees to build a new, high-resolution, low-material ITS in the second long shutdown of the LHC. As a fundamental component of the ALICE upgrade concept, this detector will open the door for a set of new high precision measurements of rare probes at very low $p_{\rm T}$. Among these are the measurement of open charm mesons down to zero $p_{\rm T}$, charmed baryons, beauty via displaced $\mathrm{D^{0}\rightarrow K\pi}$ from $\mathrm{B}$ decays and low-mass di-leptons. In addition, extensive studies are ongoing for the full reconstruction of $\mathrm{B^{+} \rightarrow \overline{D}^{0}\pi^{+}(\overline{D}^{0}\rightarrow K^{+}\pi^{-})}$ and $\mathrm{\Lambda_{b} \rightarrow \Lambda_{c}^{+}\pi^{-}(\Lambda_{c}^{+}\rightarrow pK\pi)}$.


\section*{References}
\begin{thebibliography}{9}
\bibitem{ALICE} ALICE Collaboration, \it ALICE: Physics Performance Report, Volume I\rm , J.Phys. \bf{G30}\rm, 1517(2004).
\bibitem{RaaLight} ALICE Collaboration, arXiv:1208.2711v1 [hep-ex] (2012).
\bibitem{LightFlow} ALICE Collaboration, Phys.Rev.Lett. \textbf{107}, 032301 (2011), arXiv:1105.3865v2 [nucl-ex]
\bibitem{Enhancement}I. Belikov (ALICE Collaboration), arXiv:1109.4807v2 [hep-ex] (2011).
\bibitem{Raa}ALICE Collaboration, JHEP \textbf{09}, 112 (2012), arXiv:1203.2160v4 [nucl-ex].
\bibitem{deadCone}Y. Dokshitzer and D. Kharzeev, Phys. Lett. B \textbf{519}, 199 (2001), doi:10.1016/S0370-2693(01)01130-3.
\bibitem{flow}G. Luparello, these proceedings.
\bibitem{ALICEloi} ALICE Collaboration, CERN-LHCC-2012-012 (2012). 
\bibitem{ITScdr} ALICE Collaboration, CERN-LHCC-2012-013 (2012). 
\end{thebibliography}
\end{document}